\documentclass[modern]{aastex62}

\usepackage{amsmath}
\usepackage{amssymb}

\newcommand{\dd}{\mathrm{d}}
\newcommand{\diff}[2]{\frac{\dd #1}{\dd #2}}

\newcommand{\Ndraw}{N_\mathrm{draw}}
\newcommand{\Ndet}{N_\mathrm{det}}
\newcommand{\Neff}{N_\mathrm{eff}}
\newcommand{\Nobs}{N_\mathrm{obs}}
\newcommand{\pdraw}{p_\mathrm{draw}}

\begin{document}

\title{Accuracy Requirements for Empirically-Measured Selection Functions}

\author[0000-0003-1540-8562]{Will M. Farr}
\email{will.farr@stonybrook.edu}
\email{wfarr-vscholar@flatironinstitute.org}
\affiliation{Department of Physics and Astronomy, Stony Brook University, Stony Brook NY 11794, United States}
\affiliation{Center for Computational Astronomy, Flatiron Institute, New York NY 10010, United States}

\maketitle

When conducting a population analysis on a catalog of objects the effect of the
selection function must be incorporated to avoid so-called ``Malmquist bias''
\citep{Malmquist1922,Loredo2004,Mandel2018}. Suppose we have a catalog
consisting of data $d_i$, $i = 1, \ldots, \Nobs$, that constrain the parameters
$\theta_i$ of a set of $\Nobs$ objects.  We wish infer the population
distribution function
\begin{equation}
  \diff{N}{\theta}\left( \lambda \right),
\end{equation}
which can depend on some population-level parameters $\lambda$.  The joint
posterior for the object-level parameters $\theta_i$ and population-level
parameters is \citep{Loredo2004,Mandel2018}
\begin{equation}
  \label{eq:posterior}
\pi \propto \prod_{i=1}^{\Nobs} \left[ p\left( d_i \mid \theta_i \right) \diff{N}{\theta_i}\left( \lambda \right) \right] \exp\left[ - \Lambda\left( \lambda \right) \right] p\left( \lambda \right).
\end{equation}
$p\left( d \mid \theta\right)$ is the likelihood function that describes the
measurement process for the catalog, $p\left( \lambda \right)$ is a prior, and
$\Lambda$ is the expected number of detections:
\begin{equation}
  \label{eq:selection-integral-unnorm}
  \Lambda\left( \lambda \right) \equiv \int_{\left\{ d \mid f(d) > 0 \right\}} \dd d \, \dd \theta \, \diff{N}{\theta}\left( \lambda \right) p\left( d \mid \theta \right).
\end{equation}
$f$ represents the selection function; an observation will be included in the
catalog if and only if it generates data such that $f(d) > 0$.  We factor an
overall normalization out of the population distribution so that
\begin{equation}
  \diff{N}{\theta}\left( \lambda \right) = R \xi\left( \theta \mid \tilde{\lambda} \right),
\end{equation}
with the amplitude of $\xi$ fixed in some way; $\tilde{\lambda}$ is the set of
parameters that remain once the amplitude of the population distribution is
fixed.  In this re-parameterization, $\Lambda = R x$, where $x$ is given by
\begin{equation}
  \label{eq:selection-integral}
  x\left( \tilde{\lambda} \right) \equiv \int_{\left\{ d \mid f(d) > 0 \right\}} \dd d \, \dd \theta \, \xi\left( \theta \mid \tilde{\lambda} \right) p\left( d \mid \theta \right).
\end{equation}
If $\xi$ integrates to one over all $\theta$, then $x$ is the \emph{fraction} of
sources from a population described by $\tilde{\lambda}$ that are detectable.

In simple cases the integral in Eq.\ \eqref{eq:selection-integral} can be
evaluated analytically.  But for most realistic applications it is not possible
to analytically evaluate $f$ \citep[see
e.g.][]{Burke2015,Christainsen2015,GW150914-Rate,GW150914-Rate-Supplement,Burke2017}.
Instead, the detection efficiency must be estimated by drawing synthetic objects
from a fiducial distribution, $\pdraw\left( \theta \right)$, drawing
corresponding data from the likelihood function $p\left( d \mid \theta \right)$,
and ``injecting'' these data into the pipeline used to produce the catalog,
recording which observations are detected \citep{Tiwari2018}.  This procedure
introduces uncertainty in the estimation of the selection integral; we must have
enough draws that this uncertainty does not alter the shape of the posterior
$\pi$ very much.

Given a set of detected objects with parameters $\theta_j$, $j = 1, \ldots,
\Ndet$ generated from a total number of draws $\Ndraw$ the integral in Eq.\
\eqref{eq:selection-integral} can be estimated via
\begin{equation}
  \label{eq:simple-monte-carlo-estimate}
  x \simeq \frac{1}{\Ndraw} \sum_{j=1}^{\Ndet} \frac{\xi\left( \theta_j \mid \tilde{\lambda} \right)}{\pdraw\left( \theta_j \right)}.
\end{equation}
Under repeated samplings $x$ will follow an approximately normal distribution
\begin{equation}
    x \sim N\left( \mu, \sigma \right),
\end{equation}
with
\begin{equation}
  \mu \simeq \frac{1}{\Ndraw} \sum_{j=1}^{\Ndet} \frac{\xi\left( \theta_j \mid \tilde{\lambda} \right)}{\pdraw\left( \theta_j \right)} ,
\end{equation}
and
\begin{equation}
    \sigma^2 \equiv \frac{\mu^2}{\Neff} \simeq \frac{1}{\Ndraw^2} \sum_{i=1}^{\Ndet} \left[ \frac{\xi\left( \theta_j \mid \tilde{\lambda} \right)}{\pdraw\left( \theta_j \right)} \right]^2 - \frac{\mu^2}{\Ndraw}.
\end{equation}
We have introduced the parameter $\Neff$ that gives the \emph{effective} number
of independent draws that contribute to the estimate of $x$.

Given a particular sampling of the selection function, we should marginalize
over the uncertainty in $x$.  Eq.\ \eqref{eq:posterior} becomes
\begin{equation}
  \label{eq:posterior-integrated}
    \pi \propto \prod_{i = 1}^{\Nobs} \left[ p\left( d_i \mid \theta_i \right) \xi\left( \theta_i \mid \tilde{\lambda} \right) \right] \int \dd x \, R^{\Nobs} \exp\left[ -R x \right] N\left( x \mid \mu, \sigma\right).
\end{equation}
Integrating over $-\infty < x < \infty$, we obtain
\begin{equation}
  \label{eq:approx-marginal-x-posterior}
  \pi \propto \prod_{i = 1}^{\Nobs} \left[ p\left( d_i \mid \theta_i \right) \xi\left( \theta_i \mid \tilde{\lambda} \right) \right] R^{\Nobs} \exp\left[  \frac{R \mu \left( R\mu - 2 \Neff \right)}{2 \Neff} \right].
\end{equation}
The divergence of this expression as $R\to \infty$ reflects that the normal
approximation permits non-zero probability of $x < 0$. Eq.\
\eqref{eq:approx-marginal-x-posterior} has stationary points in $R$ at
\begin{equation}
  R = R_{\pm} = \frac{\Neff \pm \sqrt{\Neff \left( \Neff - 4 \Nobs \right)}}{2 \mu}.
\end{equation}
Provided $\Neff > 4 \Nobs$ these stationary points will occur for real, positive
$R$.  In this case, the stationary point at $R_{-}$ is a local maximum; at
$R_{+}$ we have a minimum associated with the ``unphysical'' transition to the
divergent behavior as $R\to \infty$.  We have
\begin{equation}
  R_{-} = \frac{\Nobs}{\mu} \left( 1 + \frac{\Nobs}{\Neff} + 2 \left( \frac{\Nobs}{\Neff} \right)^2 + \mathcal{O}\left( \frac{\Nobs}{\Neff} \right)^3 \right).
\end{equation}
$R = \Nobs / \mu$ is the point estimate for the detection efficiency in Eq.\
\eqref{eq:simple-monte-carlo-estimate}.  Near $R = R_{-}$ a normal approximation
holds for the posterior as a function of $R$ with $\mu_R = R_{-}$ and
\begin{equation}
  \sigma_R = \frac{\sqrt{\Nobs}}{\mu} \left( 1 + \frac{3}{2} \frac{\Nobs}{\Neff} + \frac{31}{8} \left( \frac{\Nobs}{\Neff} \right)^2 + \mathcal{O} \left( \frac{\Nobs}{\Neff} \right)^3 \right).
\end{equation}
Marginalizing the normal approximation over $R$ imposing a flat-in-log $R$ prior
gives
\begin{equation}
  \label{eq:fully-marginalized-posterior}
  \log \pi \propto \sum_{i=1}^{\Nobs} \log p\left( d_i \mid \theta_i \right) \xi\left( \theta_i \mid \tilde{\lambda} \right) - \Nobs \log \mu + \frac{3 \Nobs + \Nobs^2}{2 \Neff} + \mathcal{O} \left( \Neff \right)^{-2}.
\end{equation}
The term involving $\mu$ would appear in an analysis that ignores the rate $R$
and works entirely with population distributions
\citep{Mandel2018,Fishbach2018}; the term involving $\Neff$ is a correction to
account for the uncertainty in our estimate of the selection integral.

The uncertainty in parameters is driven by the \emph{differences} in the
log-posterior.  The $R$-dependent terms contribute to such differences through
\begin{equation}
  \label{eq:pi-accuracy}
  \Delta \log \pi = \ldots - \Nobs \left( \frac{\partial \log \mu}{\partial \tilde{\lambda}} - \frac{\Nobs}{2 \Neff} \frac{\partial \log \Neff}{\partial \tilde{\lambda}} \right) \Delta \tilde{\lambda}.
\end{equation}
Both derivatives are independent of $\Neff$, so the relative contribution of the
second term to the parameter estimates is $\mathcal{O}\left( \Nobs / \Neff
\right)$.

If $\Neff$ becomes close to $4 \Nobs$ for any relevant set of population
parameters then the posterior no longer peaks in $R$ and more injections must be
obtained for an accurate analysis.

\acknowledgments

A worked example, along with the \LaTeX{} source for this document, can be found
at \url{https://github.com/farr/SelectionAccuracy}.

\bibliography{selacc}

\end{document}